\documentstyle[12pt,aasms4]{article}
\slugcomment{{\em Revised version, submitted to the Astrophysical Journal, Letters to
the Editor}}

\begin{document}
 
\title{Dark Baryons and Rotation Curves}
\author{Andreas Burkert$^{1}$ and Joseph Silk$^{2,3}$\\
{\small $^1$Max-Planck-Institut f\"{u}r Astronomie}\\
{\small K\"onigstuhl 17, D-69117 Heidelberg, Germany}\\
{\small $^2$Institut d'Astrophysique, 75014  Paris, France}\\
{\small $^3$Department of Astronomy and Physics, and Center for Particle Astrophysics}\\
{\small University of California, Berkeley CA 97420, USA}}
\authoremail{burkert at mpia-hd.mpg.de}
 
\begin{abstract}
The best measured rotation curve for any galaxy is that of the dwarf
spiral DDO 154, which extends out to about 20 disk scale lengths.
It  provides an ideal laboratory for testing the universal density
profile prediction from high resolution numerical simulations of
hierarchical clustering in  cold dark matter-dominated
cosmological models. We find that the observed rotation curve cannot be fit 
either at small radii, as previously noted, or at large radii.
We advocate a resolution of this dilemma by postulating the existence of 
a dark spheroid of baryons amounting to several
times the mass of the observed disk component and comparable to that of
the cold dark matter halo component. Such an additional mass
component provides an excellent fit to the rotation curve provided
that the outer halo is still cold dark matter-dominated with a
density profile and mass-radius scaling relation as predicted by 
standard CDM-dominated models.  The universal existence of such dark baryonic
spheroidal components provides a natural explanation of { the universal
rotation curves observed in spiral galaxies}, may  have a similar origin and composition
to the local counterpart that has been detected as MACHOs in our own galactic halo
via gravitational microlensing, and is  consistent with, and even motivated by,
primordial nucleosynthesis estimates of the baryon fraction.
\end{abstract} 

\keywords{dark matter --- galaxies: formation and halos}

\section{Introduction}

Cosmological cold dark matter theories of
structure formation in the universe via hierarchical
merging are in some difficulty.
Recent high-resolution cosmological N-body simulations
(Navarro et al. 1996a, 1997) have substantially 
improved our understanding of the equilibrium density profiles
which dark matter halos  achieve when formed through hierarchical
clustering. It has been shown that the violent, collisionless dynamical
relaxation processes during the formation phase of dark matter
halos lead to equilibrium profiles that have similar shapes,
independent of halo mass, initial density
fluctuation spectrum, and  adopted cosmological model.
All dark matter profiles can be well fit by the simple formulae for
density and mass:
\begin{eqnarray}
\rho_{DM}(r) &  = & \frac{4 \rho_0}{(r/R_s)(1 + r/R_s)^2} \nonumber \\
M_{DM}(r) & = & M_0 \times [ \ln (1+r/R_s) - (r/R_s)/(1+r/R_s) ] ,
\end{eqnarray}
\noindent where $\rho_0$ is the density of the dark matter halo evaluated at the
scale radius $R_s$ and $M_0$ is the characteristic mass; $R_s$ and $M_0$
(or $\rho_0$) are free parameters. $M_0$ is a function of the total virial mass
$M_{200}$ inside the virial radius $R_{200}$
\begin{equation}
M_0 = \frac{M_{200}}{\ln (1+R_{200}/R_s) - (R_{200}/R_s)/
(1 + R_{200}/R_s)}
\end{equation}
\noindent where $R_{200}$ denotes the radius inside which the averaged 
overdensity of dark matter is 200 times the critical density of the
universe. 
The simulations also show that, for any particular cosmology, 
$R_s$ and $M_0$ are strongly correlated:
%\begin{equation}
$R_s = 1.63 \times 10^{-2-c} (M_{200}/M_{\odot})^{1/3} h^{-2/3}\rm \, kpc$
%\end{equation}
%\noindent 
where $c = \log (R_{200}/R_s)$ is approximately 1.4 for 
low-mass dark matter halos as considered in this paper. 
Low-mass halos are denser than more massive systems.
This results from the fact that lower mass halos form earlier,
at times when the universe is significantly denser. Dark matter halos therefore
represent a 1-parameter family, being completely described by equation 1
and their virial mass $M_{200}$ or virial radius $R_{200}$.
The universal profile has been verified by
simulations to halo masses as small as $M_{200} \approx 10^{11}
M_{\odot}$ but there is no reason to believe that these results would not
be valid for halos which are even one order of magnitude lower in mass.

Substantial progress has also been made { within the past decade }
on the observational front. The observations show that the
rotation curves of low-luminosity disk galaxies and low-surface 
brightness galaxies are strongly
dark matter-dominated (Persic \& Salucci 1988, 1990, Puche \& Carignan 1991,
Broeils 1992, Persic et al. 1996, de Blok et al. 1996).
The spherically-averaged mass distributions
$M(r)=V_c^2 r/G$ can be derived from the measured circular
velocity distributions $V_c(r)$
of  gaseous galactic disks. Subtraction of the contributions by the
visible components gives the dark matter mass profiles $M_{DM}(r)$ or
the corresponding dark matter rotation curves $V_{c,DM}(r) =
(G M_{DM}/r)^{1/2}$.
Given these theoretical and observational 
developments, an important question arises as to  whether
the theoretically predicted rotation curves, determined by equation 1,
are in agreement with the observations. 

\section{The dark matter halo of DDO 154}

The low-surface brightness dwarf galaxy DDO 154 has one of the most extended and best-studied dark matter halo rotation curves
(Carignan \& Freeman 1988, Carignan \& Beaulieu 1989, Carignan \& Purton 1997), with a precise 
decomposition into contributions from stars, gas and dark matter. 
It is also one of the most gas-rich systems known with an inner stellar
disk component of  mass  $5(\pm 2.5) \times 10^7 M_{\odot}$ and an 
extended HI disk with scale radius of $0.4(\pm 0.05)$ kpc and  
mass of approximately $3(\pm 1) \times 10^8 M_{\odot}$ { (Carignan \& Beaulieu 1989). }
The shape of the rotation curve, even in the innermost regions, 
is completely dominated by  the dark matter halo with a total mass
{ within $R_{200}$} of about
$5 \times 10^9 M_{\odot}$ { (see section 3).}

DDO 154 is a { good} candidate with which to test cosmological models.
Fig. 1 (upper panel) shows the  dark matter rotation curve of 
DDO 154 and compares it with
the profiles  predicted from cosmological models.
The stellar disk and the HI contribution (Carignan \& Freeman 1988)
have been subtracted, adopting a stellar mass-to-light ratio
of (M/L$\rm_B$)$_*$ = 1.
The error bars indicate the uncertainties in the observations.
Note that the data extends out to 21 disk scale lengths and 
that the rotation curve clearly decreases beyond 5 kpc.
The thick dashed and
dotted curves fit the theoretically predicted rotation curves 
as determined by equation 1 to the innermost and outermost regions, respectively.
Fitting the inner regions (dotted curve, labeled A) is well known to pose a problem
(Flores \& Primack 1994, Moore 1994, Burkert 1995) 
as the theoretical models predict a central $r^{-1}$ density cusp,
whereas the observed
velocity profile indicates a large isothermal core with a constant
density. As a result, the theoretical models lead to far more mass in the
innermost region than is  seen. It has been suggested that this
discrepancy could be solved by assuming secular processes 
(e.g. violent galactic winds) in the baryonic component (Navarro et al. 1996b)
which could also affect the innermost parts of dark matter halos. 

However a similar problem exists in the outermost regions.
Whereas the observed dark matter rotation curve clearly 
decreases beyond 5 kpc, the theoretically
predicted universal profile fit leads to a very massive and extended
halo ($M_{200}=1.8 \times 10^{13} h^{-1} M_{\odot};
R_{200} = 863 h^{-1}$ kpc, where $h$ is the Hubble constant in units
of 100 km s$^{-1}$ Mpc$^{-1}$) with a rotation curve that increases beyond 8 kpc.
The outer regions certainly cannot be affected by secular mass loss
involving a baryonic component, and a different explanation must be sought. 
The dashed line in the upper panel of Fig. 1 (labeled B) shows a fit to the 
outer rotation curve
of DDO 154. In this case the dark matter excess in the inner regions
is unacceptably large.Approximately  $10^9 M_{\odot}$ in dark matter 
would have to be moved from the inner 2 kpc into the region
between 2 kpc and 4 kpc to explain the discrepancy between the
observed inner rotation curve and the predicted one. This is far more
than   would be expected as a result of secular processes. 

Figure 2 compares the values of $M_0$ and $R_s$ (see equation 1), derived
from fitting the inner or outer rotation curve of DDO 154 
with the standard, cluster-normalized CDM predictions
(Navarro et al 1996a, 1997).
Note that both fits require pairs of  parameter values   which
are not in agreement with theory. In order to lie within the
expected parameter range one would have to choose values for
$M_0$ and $R_s$ which do not fit the rotation curve,
either in the inner or in the outer regime.

\section{The dark baryonic component of DDO 154}

What is wrong with either CDM or with the failure to fit the rotation curve of
DDO 154? There is accumulating evidence that the CDM models can
account for many aspects of large-scale structure and therefore
should not be dismissed. Errors in determining the rotation curve
of DDO 154 are equally unlikely due to its very regular velocity field 
from which $V_c(r)$ can be derived and unambiguously 
corrected for warping. As described above, secular processes cannot explain
the discrepancy. There also cannot exist much more mass in the
HI disk, e.g. in $H_2$, as such a massive disk would be gravitationally
unstable and would efficiently 
form molecular clouds and stars, in contradiction to the
observations.

We propose, that in addition to the CDM dark matter halo which can
be represented by the one-parameter family of universal profiles,
DDO 154 contains a second dark but compact baryonic component whose 
presence we infer in order to explain the observed rotation curve.
Because of the arguments mentioned above, this new component cannot be 
gaseous and located in the disk. We identify it with a 
{\it spheroidal distribution of massive compact baryonic objects, MACHOS},
located in the inner regions of the dark matter halo.

There is in fact ample room for such a subdominant baryonic component,
relative to the massive dark halos of the standard CDM theory
in the form of MACHOS. Primordial nucleosynthesis requires
a baryonic component of { $\Omega_b h^2 \approx 0.015 \pm 0.008$
(Kurki-Suonio, Jedamzik \& Mathews 1997, Copi, Schramm \& Turner 1995)
whereas the observed value
for stellar and gaseous components in disks lies in the range
of $\Omega_d  \approx 0.004 \pm 0.002$
(Persic \& Salucci 1992, note that $\Omega_d \approx 0.006$} for DDO 154, assuming 
$\Omega_{halo} = 0.1$).
Where are all these additional  baryons? Most of them should be found in
galaxies unless winds have strongly depleted all the initial baryons during
the protogalaxy phase. Even if protogalactic winds
have occurred, as is suggested by theory and observation, it is unlikely
that all of the baryons that are not in the disk would have been lost.
The recent MACHO experiments (Alcock et al. 1996)
demonstrate that there indeed exists a substantial baryonic component in a
more extended, spheroidal distribution in the Milky Way, at least between
the Sun and the LMC, in addition
to the observed stellar and gaseous baryonic components.
It is likely that a similar, yet hitherto unobserved, component also exists
in DDO 154.

The mass distribution $M_{sph}(r)$ of the proposed spheroidal dark baryonic
component is determined by 
\begin{equation}
M_{sph}(r) = V_{c,DM}^2 r/G - M_{DM}^*(r)
\end{equation}
\noindent which is the difference in total mass inside a radius $r$ as
expected from the dark matter rotation curve $V_{c,DM}$ and the dark matter
mass $M_{DM}^*$ inside $r$. We consider two extreme possibilities for
determining $M_{DM}^*(r)$. One possibility is just to adopt the universal
profile predicted by hierarchical infall models (equation 1): 
$M_{DM}^*(r) = M_{DM}(r)$. In this case the
baryonic dark component formed at the same time or even earlier than
the extended dark halo. In the opposite limit we imagine that the
MACHO spheroid formed after the extended dark halo profile was established.
In this case, if the dark baryonic component is of comparable mass
inside a certain radius, the dark halo within this radius will undergo
adiabatic contraction. For simplicity we will assume spherical symmetry
and a formation timescale of the MACHO halo which is long compared to
the dynamical timescale of the dark halo.
In this case the corrected dark matter mass profile is determined by
(Binney \& Tremaine 1987)
%\begin{equation}
$M_{DM}^*(r) = M_{DM}(r^*) 
$
%\end{equation}
where $M_{DM}(r)$ is determined by equation 1 and
\begin{equation}
r^* = r \times \frac{M_{DM}(r^*) + M_{sph}(r)}{M_{DM}(r^*)}.
\end{equation}
\noindent Inserting $M_{sph}$ (equation 3) into equation 4 and choosing
values for $M_0$ and $R_s$, one can determine iteratively the adiabatically
contracted dark matter mass profile $M_{DM}^*$ and $M_{sph}(r)$.
%using equation 5.

Our approach seems at first sight a very
contrived solution to the problem of the rotation curve, 
as we attribute the discrepancy
between observations and theory to a third, invisible component which
introduces an extra degree of freedom. How can this idea be tested?
There exist in fact very strong constraints on the shape of $M_{sph}(r)$,
which must be fulfilled in order for our model to be physically
correct.
First of all, the total mass distribution of the dark spheroid has to be 
positive everywhere and increase monotonically with increasing
radius, up to a maximum radius $R_{sph}$ beyond which it has to remain 
constant. Second, the density of the spheroid
$\rho_{sph}(r) = (dM_{sph}/dr) / (4 \pi r^2)$ must decrease 
monotonically with increasing radius.

These constraints require that the dark matter
rotation curve lies everywhere below the observed rotation curve.
Dark matter fits like the curves labeled A and B in the upper 
panel of Fig. 1 are ruled out. 
$R_{sph}$ could in principle lie outside  the observed radius regime.
However for DDO 154 the measured rotation curve extends out to
21 disk scale lengths. We identify the dark spheroid with
an intermediate baryonic component that presumably formed during the 
dissipative protogalactic collapse phase, and which should therefore be more 
centrally concentrated than the dark matter halo. The rotation curve
of DDO 154 clearly decreases outside 5 kpc, and this indicates
that a large fraction of its mass lies at smaller radii. As the 
dark matter halo dominates the total mass in the outer regions, we 
expect that the mass distribution of the baryonic spheroid 
becomes constant and equal to its total mass $M_{sph}^{tot}$ 
at a radius $R_{sph} \leq 5 $ kpc.

Note that the radial dependence of $V_c$ and $M_{DM}$
is given by the observations and equation 1. It is therefore not trivial
that  the difference of the terms on the right hand side of equation 3
gives a profile which meets all of the constraints mentioned above,
even if the dark matter scale parameters $M_0$ and $R_s$ are treated
as free parameters. We have indeed found such a solution,
however only for very special values of $M_0$ and $R_s$, values which 
for the models with and without adiabatic contraction lie within the
small shaded and dark regions of figure 2, respectively.  
These sets of solutions uniquely determine the
total mass of the spheroid $M_{sph}^{tot}$, and hence, via equation 2,
the total mass $M_{200}$ of the dark matter halo.

The solid lines in the upper and lower panels of figure 1 provide an 
 excellent fit to the rotation curve,
adopting our Macho model with the non-baryonic dark matter halo parameters
fixed at $M_0 = 4 \times 10^9 M_{\odot}$ and $R_s = 4.5$ kpc
for the case without adiabatic contraction and
$M_0 = 3 \times 10^9 M_{\odot}$ and $R_s = 5.5$ kpc
if adiabatic contraction is included.
The inserts within  Figure 2 show the mass and density profiles 
of the dark baryonic spheroid and the dark matter halo without and with
adiabatic contraction. Note that in the former case without adiabatic contraction, the MACHO
component would dominate the inner mass and density distribution, whereas
in the latter case the baryonic and non-baryonic halos have comparable masses
in the inner region. For
all acceptable models we find $R_{sph} \approx 5$ kpc,
$M_{sph}^{tot} \approx 1.5 \times 10^9 M_{\odot}$ 
and $M_{200} \approx 5 \times 10^9 M_{\odot}$. The
density profiles of the  MACHO component can be well approximated by
an isothermal sphere with a constant velocity dispersion of $\sigma = 40$ km/s.

There exists an additional independent constraint in order for our model 
to be acceptable
within the framework of standard cosmology: $M_0$ and $R_s$ must
follow the tight relationship, predicted by cosmological models.
Indeed, figure 2 shows that the values which have to be adopted in order to give a
physically correct mass distribution overlap with the parameter space  of values
expected from standard cosmology, in contrast to the
one-component dark matter fits (starred points).
Moreover, as we identify the dark spheroid with
the missing baryonic component, early universe nucleosynthesis
requires that $\Omega_b/\Omega_d = (M_{sph}^{tot}+M_{visible}^{tot})/
M_{visible}^{tot} \approx  5 (\pm 3)\times h^{-2}$,
where $M_{visible}^{tot} = 3 \times 10^8 M_{\odot}$ is the total visible mass
of DDO 154. For the areas of possible solutions we
find a dark-to-luminous mass fraction of approximately 6, in agreement with the
expectations.

Note that an arbitrary choice of $M_0$, $R_s$ and $M_{sph}$ would almost certainly 
fail to meet all of these constraints. The excellent agreement of our model
with the predictions from cosmological models of structure formation and
primordial nucleosynthesis provides additional evidence for the presence of a
dark baryonic component in DDO 154.

\section{Discussion and Conclusions}
 
Our three component model of luminous baryons in a disk configuration, and
MACHOS and cold dark matter in a spheroidal distribution,
can reconcile the most detailed observations of a rotation curve to
date with the hierarchical clustering theory of galaxy formation.
This might be the first (indirect) detection of a MACHO component 
in another galaxy.
It also has allowed us to study in detail for the first time the internal
density distribution of a dark baryonic spheroid due to 
the excellent high-resolution data of DDO 154's rotation curve.

The structure of the MACHO spheroid in 
DDO 154 is surprisingly similar to the MACHO halo of the Milky Way,
the existence and mass of which has been inferred from a completely different
method: gravitational microlensing events of stars in the Large Magellanic Cloud.
Alcock et al. (1996, 1997) find for the Galaxy that within 50 kpc (14 disk scale lengths) 
the total masses of MACHOS and dark matter are comparable and of order
$2.5 \times 10^{11} M_{\odot}$. This is 4 to 5 times the mass of the galactic
disk ($M_{d} \approx 6 \times 10^{10} M_{\odot}$). The dark 
baryonic spheroid of DDO 154 also
extends out to 14 disk scale lengths at which radius the mass of the MACHO halo
is again similar to the mass of the dark matter halo, with, in this
case, a mass of order $1.5 \times 10^9 M_{\odot}$. The inferred MACHO mass is
also of order 5 times the mass of the HI disk.
This agreement in the relative mixture of dark matter,
MACHOS and disk material indicates that these components formed
in both galaxies from a similar continuous dynamical process, with 
the MACHO spheroid representing a presumably
dissipative component intermediate between the collisionless
non-baryonic dark halo and the strongly collisional, dissipation-dominated,
rotationally-supported disk. This might provide
an explanation { for the puzzling observational result that
disk galaxies have universal rotation curves
(Casertano \& van Gorkom 1991, Rubin et al. 1985, Persic et al. 1996), 
requiring a connection between their
galactic disks and their dark spheroidal components. Universal rotation 
curves would be expected}
for galaxies of any mass  where the relative mass and radius ratios between
the dark matter halo, the MACHO halo and the disk are universal numbers.

{ Our results indicate that the baryonic component of DDO154 and probably
also of other disk galaxies consists of two components, a spheroidal
MACHO component which represents about 22\% of the total mass and 
a visible disk component with only 4\% of the total mass. 
The total baryon fraction in galaxies is then of the order of a quarter of the
total mass, a value which is higher than expected from the primordial
nucleosynthesis predictions if the relative mixture of
baryonic and non-baryonic matter is universal.  This baryon segregation
could result from dissipative processes during the formation of halos 
which concentrate the baryons relative to the nondissipative dark matter.

The origin of two separate baryonic components, namely a dominant dark
spheroidal component and a disk component,
is an interesting and yet unsolved theoretical puzzle
which could provide important information on the dissipative formation 
history and evolution of galaxies.}

\acknowledgments
 
We thank Dr. C. Carignan for making his new data of DDO 154's rotation curve
available prior to publication, Dr. J. Navarro for sending us a subroutine
that generates the scaling relations as predicted from cosmological models
and Dr. S. White and the referee for helpful suggestions.
The research of J.S. has been supported in part by grants from
NASA and NSF, and he also  acknowledges with gratitude the hospitality
of the Institut d'Astrophysique de Paris  as a  Blaise-Pascale
 Visiting Professor, and the Institute of Astronomy
at Cambridge as a Sackler Visiting Astronomer.

\clearpage
 
\begin{center}
{\small FIGURE CAPTIONS}
\end{center}
 
\vspace{1.0cm}
 
{\normalsize F{\footnotesize IG}.}
1. {\it Upper panel}: The dark matter rotation curve of DDO 154 is 
shown with error bars.
The dotted line (A) shows a fit to the inner parts  of the rotation
curve, adopting the dark matter halo structure as predicted by
cosmological models. The dashed line (B) shows a dark matter halo 
fit to the outer part 
of the rotation curve. The solid line  shows the fit achieved
with the 2-component MACHO model without adiabatic contraction,
assuming dark matter halo parameters
$M_0 = 4 \times 10^9 M_{\odot}$ and $R_s = 4.5$ kpc. The lower
dashed and dot-dashed curves show the contribution to the rotation curve of
the MACHO spheroid and the dark matter halo (C), respectively.

{\it Lower panel}: The 2-component MACHO model with adiabatic contraction
is shown, adopting halo parameters $M_0 = 3 \times 10^9 M_{\odot}$ and 
$R_s = 5.5$ kpc. The dashed curve shows the contribution by
the MACHO spheroid. The dot-dashed curve (D) and the dotted curve (C$^*$)
show the contribution of the dark matter halo after and before adiabatic
contraction, respectively.
 
\vspace{1.0cm}
 
{\normalsize F{\footnotesize IG}.}
2.---  Standard cluster normalized cold dark matter models predict
that the dark matter halo scale radii $R_s$ and scale masses $M_0$
should lie within the narrow band enclosed by the two parallel solid
lines. The parallel dashed lines enclose the region of scale parameters,
expected for the less favoured COBE-normalized cold dark matter model.
One-component dark matter fits to the rotation curve of DDO 154 would
result in scale parameters as shown by the two stars for the 
inner (A) and outer (B) fits. The two-component MACHO model without
adiabatic contraction and with adiabatic contraction
requires the scale parameters to lie within the dark area (labeled C) and
the shaded area (labeled C$^*$), respectively.  The upper inserts
show the mass and density distribution ($\rho$ in 
units of $M_{\odot}$ pc$^{-3}$) for the standard model
($M_0 = 4 \times 10^9 M_{\odot}, R_s = 4.5$ kpc, star inside dark area) 
without adiabatic contraction, with the solid
and dashed lines representing the dark baryonic spheroid and the
dark matter halo, respectively.
The lower insert shows the mass distribution of the standard model
($M_0 = 3 \times 10^9 M_{\odot}, R_s = 5.5$ kpc, star inside shaded area) 
with adiabatic
contraction. The dots represent the total dark matter mass profile
as predicted from the rotation curve. The lower solid line shows
the mass profile of the MACHO halo. The dot-dashed and the dotted curves
show the mass distribution of the non-baryonic dark matter halo after and
before adiabatic contraction, respectively.

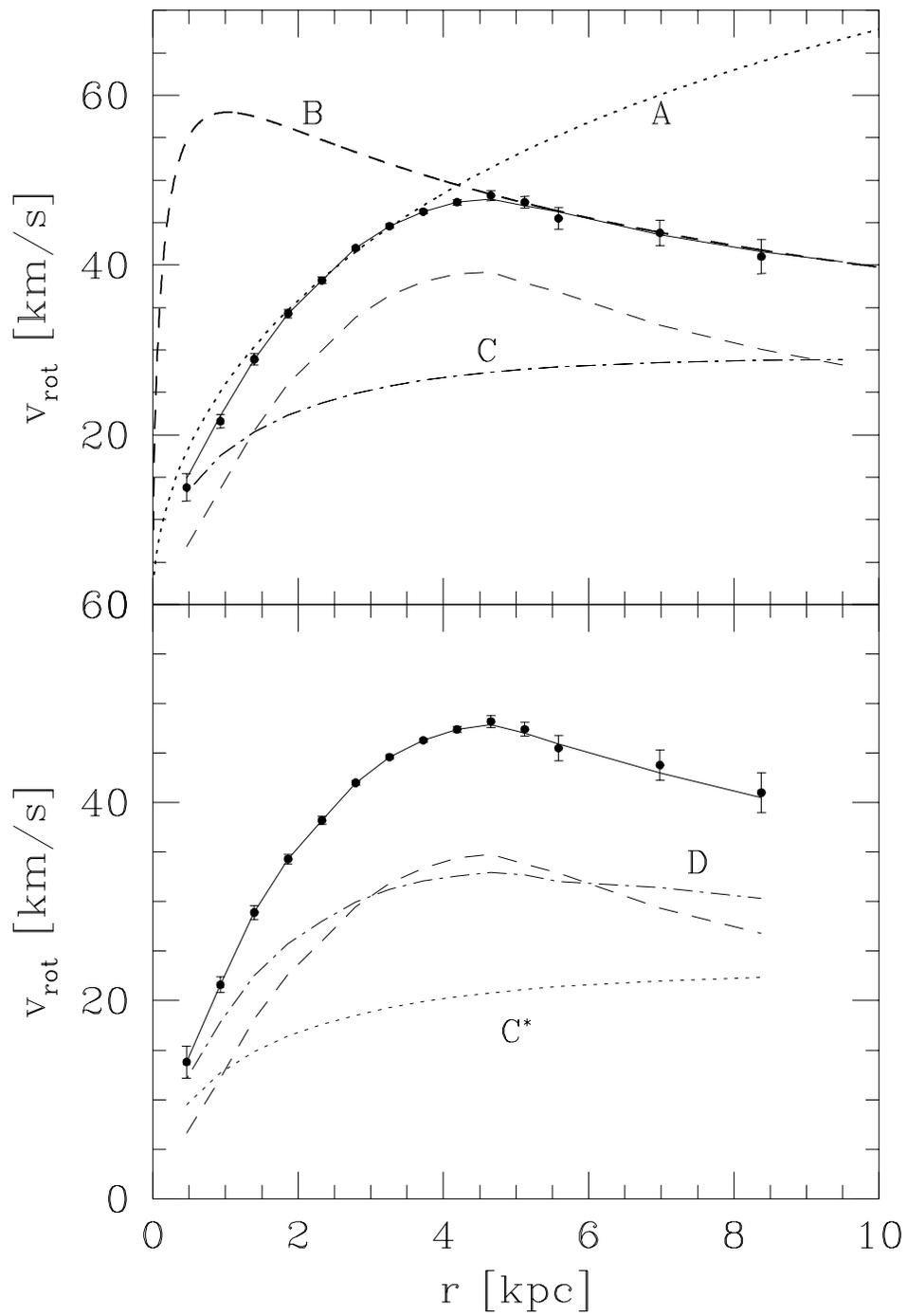
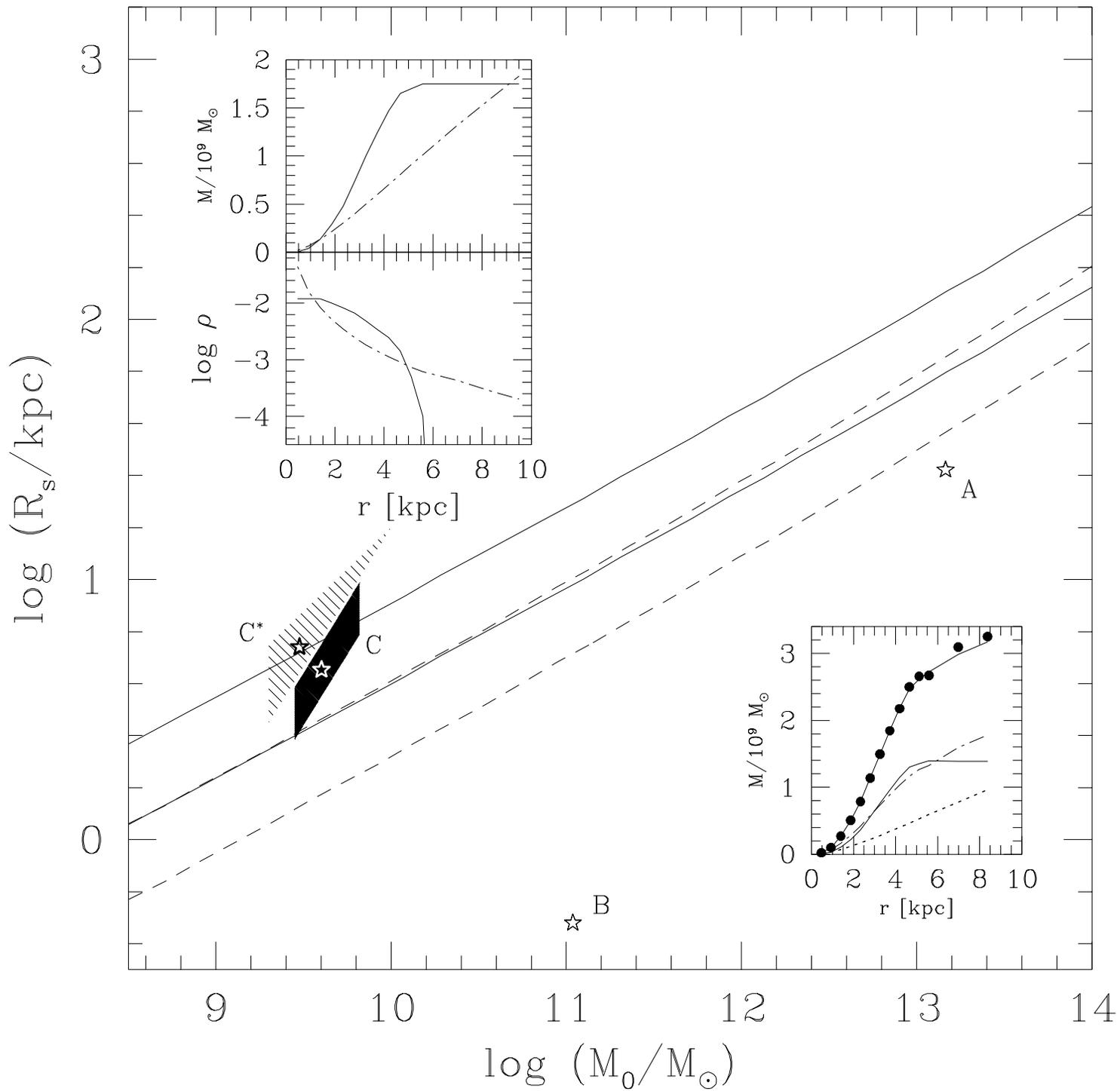
\end{document}